\documentclass[aps,prl,twocolumn,superscriptaddress,preprintnumbers,showpacs]{revtex4}
\usepackage{graphicx,color}
\begin{document}

\title
{$K\to \pi\gamma$ decays and space-time noncommutativity}

\author{B. Meli\'{c}}
\affiliation{Theoretical Physics Division, Rudjer Bo\v skovi\' c Institute, 
Zagreb, Croatia}
\author{K.Passek-Kumeri\v cki}
\affiliation{Theoretical Physics Division, Rudjer Bo\v skovi\' c Institute, 
Zagreb, Croatia}
\author{J.Trampeti\'{c}}
\affiliation{Theoretical Physics Division, Rudjer Bo\v skovi\' c Institute, 
Zagreb, Croatia}
\affiliation{Max Planck Institut f\"{u}r Physik, 
M\"{u}nchen, Germany}


\begin{abstract}
We propose the $K \rightarrow \pi\gamma$ decay mode
as a signature of the violation of the 
Lorentz invariance and the appearance
of new physics via space-time noncommutativity.
\end{abstract}

\pacs{12.38.-t, 12.39Dc, 12.39.-x, 14.20-c}
\maketitle

In this proposal we assume space-time noncommutativity (NC)
to compute the $K \rightarrow \pi\gamma$ decay forbidden
by the Lorentz invariance. 

The dynamics of the standard model (SM) forbidden flavor-changing weak decays is
described in the framework of the noncommutative SM (NCSM), where the content of particles and 
symmetries is the same as in the usual commutative space-time. 
Gauge symmetry is included in an infinitesimal form, thus SU(N) gauge symmetry is implementable.
All NC fields are expressed in terms of the usual fields via the Seiberg-Witten (SW) map  
by expansion up to first order in the NC parameter $\theta^{\mu\nu}$. 
As in other particle physics models on noncommutative space-time, a
general feature of the NCSM action is the violation of space-time
symmetries, in particular of angular momentum conservation and
discrete symmetries like P, CP, and possibly even CPT. This symmetry
breaking is spontaneous in the sense that it is broken with respect to a
fixed noncommutative background. 
 
The method for implementing non-Abelian SU(N) theories on
noncommutative space-time, based on the Seiberg-Witten map \cite{Seiberg:1999vs}, has been proposed in
\cite{Jurco:2001rq}. 
In \cite{Calmet:2001na,Behr:2002wx,Melic:2005fm,Melic:2005} this method has
been applied to the standard model
of particle physics resulting
in the noncommutative extension of the SM, called NCSM action, with the same structure 
group SU(3)$_c \times$ SU(2)$_L \times$ U(1)$_Y$ and with the same
fields and number of coupling parameters
as in the original SM.
It represents a $\theta^{\mu\nu}$--expanded effective action 
\begin{equation}
S_{\mbox{\tiny NCSM}}= S_{\mbox{\tiny fermions}} + S_{\mbox{\tiny gauge}} + S_{\mbox{\tiny{Higgs}}} +
S_{\mbox{\tiny{Yukawa}}}
\, ,
\label{eq:Sncsm}
\end{equation}
valid at very short distances, that leads to an anomaly free theory \cite{Brandt:2003fx}.
For expressions of particular contributions we refer to \cite{Melic:2005fm,Melic:2005}.
The $\theta^{\mu\nu} ={c^{\mu\nu}}/{\Lambda_{\rm NC}^2}$ is the constant, 
antisymmetric tensor,
where $c^{\mu\nu}$ are dimensionless coefficients 
of order unity and $\Lambda_{\rm NC}$ is the scale of noncommutativity.
The matter sector of the action (\ref{eq:Sncsm}), relevant to this work, 
is not affected by the freedom of choosing traces in the gauge kinetic part; 
the quark-gauge boson interactions remain the same \cite{Melic:2005fm,Melic:2005}. 
The above action is
symmetric under ordinary gauge transformations 
in addition to noncommutative ones.   

An alternative NCSM proposal was presented in \cite{Chaichian:2001py}.

Signatures of noncommutativity have been discussed within
collider physics~\cite{Hewett:2000zp,Hewett:2001im,Kamoshita:2002wq,Ohl:2004tn}, SM forbidden decays
\cite{Behr:2002wx,Schupp:2004dz,Caravati:2002ax,Devoto:2004qv,Melic:2005hb}, 
neutrino astrophysics \cite{Schupp:2002up,Minkowski:2003jg}, 
in \cite{Calmet:2004dn}, as 
well as for low-energy non-accelerator experiments
\cite{Anisimov:2001zc,Carroll:2001ws,Hinchliffe:2001im,Hinchliffe:2002km}.   

This paper represents an estimate of 
the $K \to \pi \gamma$ decay branching ratio,
based on the complete analysis of the $S_{\mbox{\tiny NCSM}}$ 
action presented in \cite{Melic:2005fm}. 
From $S_{\mbox{\tiny{Higgs}}}$  
we find the contribution proportional to the 
$M_W^2$ \cite{Melic:2005fm} 
for the $\theta$--correction to the SM vertex 
$A_{\mu}(q)W^-_{\nu}(p)W^+_{\rho}(k)$:
\begin{eqnarray}
\hspace{-.5cm}
V^{\mu\nu\rho}_{\gamma W^- W^+}&=&
i e\;\Big[g^{\mu\nu}(q-p)^\rho
+g^{\nu\rho}(p-k)^\mu +
\nonumber\\
&+&
 g^{\rho\mu}(k-q)^\nu
 +\frac{i}{2} M_W^2 \Big(
\theta^{\mu \nu} q^\rho +
\theta^{\mu \rho} q^\nu
\nonumber
\\
&+&
 \,g^{\mu\nu}(\theta q)^\rho
-g^{\nu\rho}(\theta q)^\mu + g^{\rho\mu}(\theta q)^\nu \Big)\Big]\,,
\label{2Wgamma}
\end{eqnarray}
while explicit expressions  containing important Yukawa terms for 
${\bar q}^{(i)}q^{(j)}\gamma$ and ${\bar q}^{(i)}q^{(j)}\gamma W^+$ vertices 
are given by Eqs. (76), (78) and (86) of Ref. \cite{Melic:2005fm}.
There is also an additional contribution to the vertex (\ref{2Wgamma})
from Eq. (89) in \cite{Melic:2005fm}, 
but owing to the symmetry this term vanishes. 
\begin{figure}
\begin{center}
\includegraphics[scale=0.3]{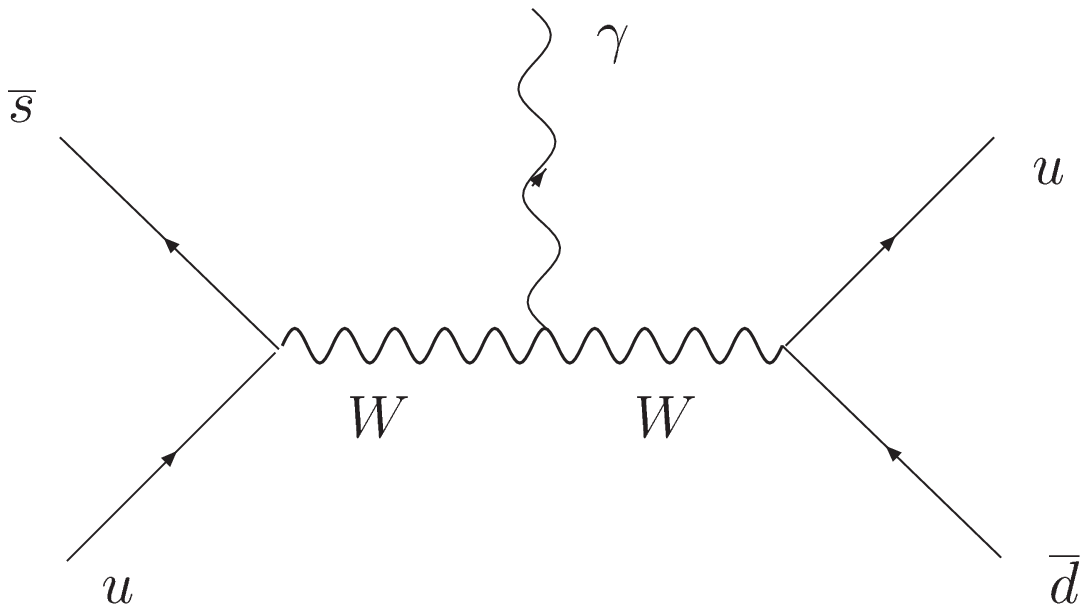}\\
\vspace*{-0.2cm} (a)
\\
\vspace*{0.5cm}
\includegraphics[scale=0.3]{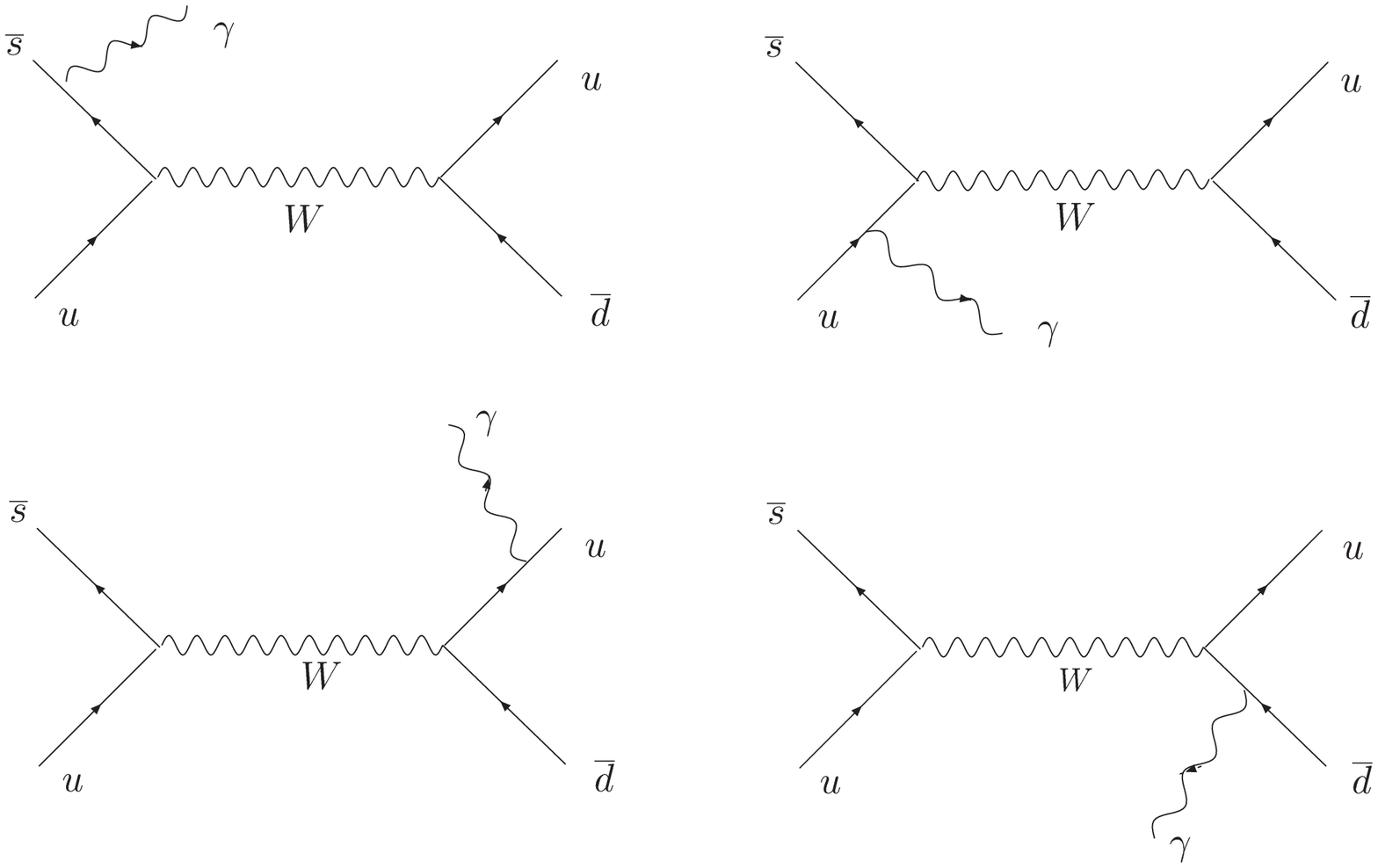}\\
\vspace*{-0.3cm} (b)
\\
\vspace*{0.5cm}
\includegraphics[scale=0.3]{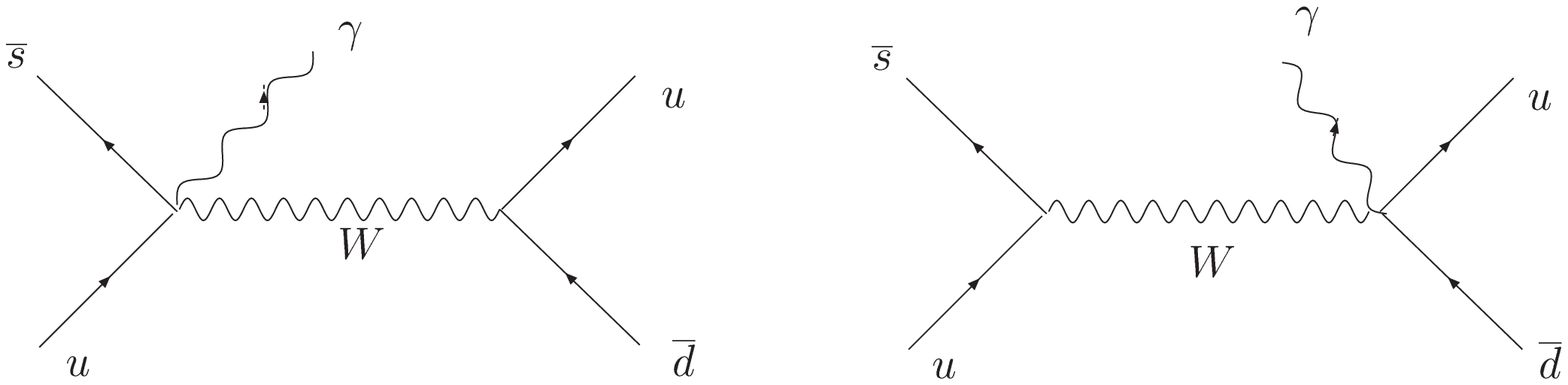}\\
\vspace*{-0.1cm} (c)
\end{center}
\caption{Feynman diagrams contributing to the free quark amplitude 
$\cal M$ responsible for the $K^+\to \pi^+ \gamma$ decay.}
\label{fig1}
\end{figure}

The free quark amplitude ${\cal M}$ contributing to the $K^+\to \pi^+\gamma$ decay arises 
from the Feynman diagrams displayed in Fig. \ref{fig1} and is given as
\begin{eqnarray}
{\cal M} 
=({\cal M}^{\rm SM}_{(a+b)} 
+ {\cal M}^{\theta}_{(a+b+c)})_{\mu}\;\varepsilon^{\mu}(q)
\,.
\label{AF}
\end{eqnarray}

The hadronic matrix element $\langle \pi^+(p)|{\cal M}|K^+(k)\rangle$ 
responsible for $K^+\to \pi^+\gamma$ decay, contains the
4-quark-current$\times$current operators from Fig. \ref{fig1}, 
and in fact represents nonperturbative quantity which
has been computed by using the vacuum saturation approximation and the 
partial conservation of the axial-vector current (PCAC):
\begin{eqnarray}
\langle \pi^+(p)|{\bar u}{\gamma_{\mu}}\gamma_5 d|0\rangle =-ip_{\mu}f_{\pi}\,.
\label{10}
\end{eqnarray}
In this way we have hadronized 
free quarks into pseudoscalar $\pi^+$- and $K^+$-meson bound states. 
Since $\langle \pi^+(p)|({\cal M}^{\rm SM}_{(a+b)})_{\mu}|K^+(k)\rangle \,\varepsilon^{\mu}(q)=0$, 
the Lorentz invariance is satisfied for the $K^+ \to \pi^+ \gamma$ process
computed in the SM, as it should be.

We obtain the following $K^+\rightarrow\pi^+ \gamma$ decay amplitude:
\begin{eqnarray}
{\cal A}^{\theta}(K^+\rightarrow\pi^+ \gamma)&=&
\langle \pi^+(p)|({\cal M}^{\theta}_{(a+b+c)})_{\mu}|K^+(k)\rangle \,\varepsilon^{\mu}(q)
\nonumber\\
&=& 
i\kappa \left({\cal A}^{\theta}_{(a+b+c)}
\right)_{\mu}
\varepsilon^{\mu}(q)\,,
\label{14}
\end{eqnarray}
where $\kappa$ is a dimensionless constant
\begin{eqnarray} 
\kappa=\frac{e\,G_F}{4\sqrt 2}\,
V^{}_{ud}V^{\dagger}_{us} f_{\pi}f_K \,,
\end{eqnarray}
while particular contributions originating from diagrams 
(a), (b), and (c) in Fig. \ref{fig1} are
\begin{eqnarray}
\Big({\cal A}^{\theta}_{(a)}\Big)_{\mu}&=&
2(k^2 (\theta {p})_{\mu}
-p^2 (\theta {k})_{\mu}-2(q\theta k)k_{\mu})\,,
\nonumber\\
\left({\cal A}^{\theta}_{(b)}\right)_{\mu}&=&
\frac{kp}{kq}\Big[(Q_u+Q_s)\Big((q\theta k)k_{\mu} -(kq)(\theta {k})_{\mu}\Big)
\nonumber\\
&&\hspace{.3cm}-\,
(Q_u+Q_d)\Big((q\theta k)p_{\mu} -(kq)(\theta {p})_{\mu}\Big)\Big]
\nonumber\\
&&\hspace{-2cm}-\,R_{\pi}\;\Big[(Q_u-Q_s)(kq)(\theta {p})_{\mu}
-i(Q_u+Q_s)\epsilon_{\mu\nu\rho\tau}q^{\nu}
(\theta {p})^{\rho}k^{\tau}\Big]
\nonumber\\
&&\hspace{-2cm}+\,R_K \Big[(Q_u-Q_d)(kq)(\theta {k})_{\mu}
+i(Q_u+Q_d)\epsilon_{\mu\nu\rho\tau}q^{\nu}
(\theta {k})^{\rho}p^{\tau}\Big]\,,
\nonumber\\
\left({\cal A}^{\theta}_{(c)}\right)_{\mu}&=&
(Q_u+Q_d)\left(p^2 (\theta {k})_{\mu}
-(kp)(\theta {p})_{\mu}+(q\theta k)p_{\mu}\right)
\nonumber\\
&+&2(m_d Q_u+m_u Q_d)\frac{p^2(\theta {k})_{\mu}}{m_d+m_u}
\nonumber\\
&-&(Q_u+Q_s)\left(k^2(\theta {p})_{\mu}
-(kp)(\theta {k})_{\mu}-(q\theta k)k_{\mu}\right)
\nonumber\\
&-&2(m_s Q_u+m_u Q_s)\frac{k^2(\theta {p})_{\mu}}{m_s+m_u}\,,
\label{17}
\end{eqnarray}
with $Q_u=2/3\,,Q_d=Q_s=-1/3$, and 
with the following kinematics and notation: 
$k=p+q$, $k^2=m_K^2,\, p^2=m_{\pi}^2,\, q^2=0$, 
$({\theta k})_{\mu}=\theta_{\mu\nu}k^{\nu}$, and  
$q\theta k = q_{\mu} \theta^{\mu\nu} k_{\nu}$\,.
The mass ratios $R_{\pi}$ and $R_K$, 
\begin{eqnarray} 
R_{\pi}&=&\frac{p^2}{kq}\frac{m_d-m_u}{m_d+m_u}\,,\;\;
R_K=\frac{k^2}{kq}\frac{m_s-m_u}{m_s+m_u}\,,
\label{20} 
\end{eqnarray}
are evaluated for 
$(m_d-m_u)/(m_d+m_u) \simeq 1/3.5$ and  $(m_s-m_u)/(m_s+m_u) \simeq 1$ 
\cite{Jamin:2001zr}\,. However, due to the numerical insignificance,
in the following we neglect all terms proportional to $R_{\pi}$.

Certain contributions to the amplitudes (\ref{17}) from the Yukawa parts of 
(b) and (c) classes of Feynman diagrams in Fig. \ref{fig1} 
combine through the ''charge - mass`` interplay 
in such a way as to manifest the SU(2) and SU(3) 
symmetry breaking via $m_d-m_u$ and $m_s-m_u$ mass differences and, more important,
to maintain the classical gauge invariance of the total amplitude (\ref{14}). 

Taking the kaon at rest and performing the phase-space integrations
we find the following rate:
\begin{eqnarray}
&&\hspace{-.9cm}BR(K^+\rightarrow\pi^+ \gamma)
\nonumber\\
&&
=\tau_{K^+}\Gamma(K^+\rightarrow\pi^+ \gamma)
\nonumber\\
&& \simeq \tau_{K^+}\frac{\alpha}{128} G^2_F f^2_{\pi}f^2_K|V^{}_{ud}V^{\dagger}_{us}|^2 
\frac{m_K^5}{\Lambda^4_{\rm NC}}\Big(1-\frac{m^2_{\pi}}{m^2_K}\Big)
\nonumber\\
&& \hspace{.3cm}\times \Big[\,1-\frac{50}{27}\,\frac{m^2_{\pi}}{m^2_K}
+\frac{25}{27}\,\frac{m^4_{\pi}}{m^4_K}\Big] 
\nonumber\\
&& \simeq
0.8 \times10^{-16}\;
\left(1\; {\rm TeV}/\Lambda_{\rm NC}\right)^4\,,
\label{16}
\end{eqnarray}
where $\tau_{K^+}$ is the $K^+$ meson mean life. 

Considering the $K^+\to \pi^+\gamma$ experiment we report on the
Brookhaven collaboration E949 who recently published a new upper limit on the branching ratio
$BR(K^+\rightarrow\pi^+ \gamma)<2.3 \times 10^{-9}$, at 90\%CL \cite{Artamonov:2005ru}.
This result is based on the data analysis primarily used 
to extract a $K^+\to \pi^+\gamma\gamma$ result near the $\pi^+$ kinematic endpoint
to test unitarity corrections of the 
chiral perturbation theory. 
Having the $K^+\to \pi^+\gamma\gamma$ background under control, the  
limit achieved on the $K^+\to \pi^+\gamma$ branching ratio is 
about 150 times better \cite{Artamonov:2005ru} 
with respect to the previous results of the E787 collaboration \cite{Eidelman:2004wy}.

Here we have presented
theoretical computation of the $K^+\rightarrow\pi^+ \gamma$ branching ratio in the NCSM.
The inclusion of the NC parts of
the classes (a) and (b) of Feynman diagrams (Fig. \ref{fig1})
into the total amplitude ${\cal A}^{\theta}_{(a+b+c)}$  
represents the novel feature in comparison with previous estimate \cite{Trampetic:2002eb} 
(see also the relevant Feynman rules in Ref. \cite{Melic:2005fm}).
Main enhancement of the rate (\ref{16}) with respect to result of Ref. \cite{Trampetic:2002eb} 
is coming from the class (a) of Feynman diagrams in Fig. \ref{fig1} 
via $\theta$--correction to the SM vertex $A_{\mu}W^-_{\nu}W^+_{\rho}$
(\ref{2Wgamma}) discovered through analysis of the $\theta$-expanded Higgs sector action
$S_{\mbox{\tiny{Higgs}}}$ in \cite{Melic:2005fm}. 
The rate (\ref{16}) is about an order of magnitude higher with respect to the first, 
incomplete estimate \cite{Trampetic:2002eb} based only on the gauge invariant part 
of the amplitude ${\cal A}^{\theta}_{(c)}$ from (\ref{17}).
Our prediction is correct within the approximation made, 
i.e., by neglecting hardly controllable corrections ($1/N_c$, etc.) to 
the vacuum saturation approximation and to the PCAC.

In the framework of NCSM, i.e. the minimal NC extension of SM,
another SM forbidden decay was recently examined,
quarkonia $\rightarrow  \gamma \gamma$ \cite{Melic:2005hb}.
Although the SM forbidden decays of this kind could serve as
a potentially good place for the discovery of space-time noncommutativity,
the existing experimental limits are too weak to set a meaningful
bound on $\Lambda_{\rm NC}$
(for other estimates from the literature,
see \cite{Melic:2005hb} and references therein).
Collider scattering experiments could offer another
``laboratory'', also very sensitive to the space-time NC signals.
The first limits on noncommutative QED from an
$e^+ e^-$ collider experiment, yielding
$\Lambda_{\rm NC} > 141$ GeV at $95\%$ confidence level,
was obtained by the OPAL collaboration \cite{Abbiendi:2003wv}.
The high precision of the future linear colliders could enable
searches for noncommutativity by measuring deviations from the SM
polarization observables \cite{Hewett:2000zp,Hewett:2001im,Kamoshita:2002wq,Ohl:2004tn}.
In such a way a bound on NC parameters could be set
more restrictively,
since the near-future collider experiments will be
sensitive to energy scales corresponding to $\Lambda_{\text{NC}} \sim
1\,\rm TeV$.

To conclude, concerning the considered $K \to \pi \gamma$ decay
and the possibility that the space-time noncommutativity is observed
in such a decay,
our theoretically predicted signature
is relatively small.  
However, the arrival of new facilities should be encouraging,
because further machines are expected to yield a production
of ${\bar K}K$ pairs
that might be larger by a number of orders of magnitude 
\footnote{Applying the OPAL Collaboration 
bound $\Lambda_{\rm NC}\,>\, 141$ GeV \cite{Abbiendi:2003wv} on our result (\ref{16})
yields $BR(K^+\rightarrow\pi^+ \gamma)\,<\, 2 \times 10^{-13}$ for which we believe 
that it could be reached in the future kaon factories.}.
\\

We want to thank S. Chen, S. Kettell, and J. Wess for fruitful discussions. 
This work was supported by the
Ministry of Science, Education, and Sport of the Republic of Croatia under Contract No. 0098002.

\end{document}